# Grain-size dependent high-temperature ferromagnetism of polycrystalline Mn$_x$Si$_{1-x}$ ($x$~0.5) films

S.N. Nikolaev[1], A.S. Semisalova[2,3], V.V. Rylkov[1,4(a)], V.V. Tugushev[1,5(b)], A.V. Zenkevich[6,7], A.L. Vasiliev[1], E.M. Pashaev[1], K.Yu. Chernoglazov[1], Yu.M. Chesnokov[1], I.A. Likhachev[1], N.S. Perov[3], Yu.A. Matveyev[6,7], O.A. Novodvorskii[8], E.T. Kulatov[5], A.S. Bugaev[4,6], Y. Wang[2], S. Zhou[2]

[1] National Research Centre "Kurchatov Institute", 123182 Moscow, Russia
[2] Helmholtz-Zentrum Dresden-Rossendorf, Institute of Ion Beam Physics and Materials Research, Bautzner Landstrasse 400, 01328 Dresden, Germany
[3] Faculty of Physics, Lomonosov Moscow State University, 119991 Moscow, Russia
[4] Kotel'nikov Institute of Radio Engineering and Electronics RAS, 141190 Fryazino, Moscow Region, Russia
[5] Prokhorov General Physics Institute RAS, 119991 Moscow, Russia
[6] Moscow Institute of Physics and Technology, 141700 Dolgoprudny, Moscow Region, Russia
[7] National Research Nuclear University "MEPhI", 115409 Moscow, Russia
[8] Institute on Laser and Information Technologies RAS, 140700 Shatura, Moscow Region, Russia

___________________________________
E-mails: (a) vvrylkov@mail.ru; (b) tuvictor@mail.ru

**Abstract**

We present the results of a comprehensive study of magnetic, magneto-transport and structural properties of nonstoichiometric Mn$_x$Si$_{1-x}$ ($x$≈0.51-0.52) films grown by the Pulsed Laser Deposition (PLD) technique onto Al$_2$O$_3$(0001) single crystal substrates at $T$ = 340°C. A highlight of our PLD method is the using of non-conventional ("shadow") geometry with Kr as a scattering gas during the sample growth. It is found that studied films exhibit high-temperature (HT) ferromagnetism (FM) with the Curie temperature $T_C$ ~ 370 K accompanied by positive sign anomalous Hall effect (AHE); they also reveal the layered polycrystalline structure with a self-organizing grain size distribution. The HT FM order is originated from the bottom interfacial nanocrystalline layer, while the upper layer possesses the low temperature (LT) type of FM order with $T_C \approx$ 46 K, gives essential contribution to the magnetization at $T \leq$ 50 K and is homogeneous on the nanometer size scale. Under these conditions, AHE changes its sign from positive to negative at $T \leq$ 30K. We attribute observed properties to the synergy of self-organizing distribution of Mn$_x$Si$_{1-x}$ crystallites in size and peculiarities of defect-induced FM order in PLD grown polycrystalline Mn$_x$Si$_{1-x}$ ($x$~0.5) films.

## 1. Introduction

Mn$_x$Si$_{1-x}$ ($x$≈0.5) alloyed films with composition close to the manganese monosilicide MnSi are materials with exceptional combination of magnetic and transport properties; at the same time they are promising for spintronic applications [1-7]. The perfect single crystal ε-MnSi with B20-type of structure possesses at low temperatures (≤ 30 K) intriguing magnetic and transport phenomena caused by formation of new magnetic quasiparticles – skyrmions (see [6] and references therein). On the other hand, the Mn$_x$Si$_{1-x}$ ($x$≈0.5) thin layers grown on Si(001) or Al$_2$O$_3$(0001) substrates demonstrate the high-temperature (HT) ferromagnetism (FM) with the Curie temperature $T_c$ of the order of room temperature [2-4]. This fact is in contrast to the case of bulk $\varepsilon$-MnSi single crystal, where only the low-temperature (LT) FM was reported with $T_C \approx 30$ K [8, 9]. The HT FM order at $x \cong 0.506$ (that just corresponds to single crystal $\varepsilon$-MnSi belonging to berthollides [9, 10]) was observed in the $Mn_xSi_{1-x}/Si(001)$ structures but at enough small $Mn_xSi_{1-x}$ film thickness less than one $\varepsilon$-MnSi monolayer [2, 11]. This order is explained by the formation of c-MnSi phase with B2-like (CsCl) crystal structure stabilized with tetragonal distortion due to favorable lattice mismatch between the film and substrate [1]. Recently we reported the HT FM appearance with $T_C \approx 330$ K in 70 nm thick $Mn_xSi_{1-x}$ ($x \approx 0.52$-$0.55$) films grown on the $Al_2O_3(0001)$ substrates by pulsed laser deposition (PLD) technique [3,4]. We argued that the observed HT FM has a defect-induced nature: it is due to formation of local magnetic moments on the Si vacancies inside the MnSi matrix and the strong exchange coupling between these moments mediated by spin fluctuations of itinerant carriers [12]. The $Mn_xSi_{1-x}$ films in [3, 4] were deposited at a relatively slow deposition rate (~2 nm/min) using PLD method in a conventional "direct" geometry (DG) when the surface of $Al_2O_3(0001)$ substrate is exposed to the Mn-Si laser plume. Accordingly to atomic-force microscopy (AFM) measurements, the structure of thus grown films is mosaic, with the crystallite size ~0.5-1 μm.

In this work we present pioneering results of a comprehensive study of magnetic, magneto-transport and structural properties of the $Mn_xSi_{1-x}$ ($x \approx 0.52$) polycrystalline films grown by PLD technique employing unconventional "shadow" geometry (SG) with Kr as a buffer gas. As compared to the conventional "direct" geometry (DG) of Ref. [3, 4], in the SG method the effective scattering of ablated particles in the buffer gas results in the lower energy of the depositing atoms as well as very high deposition rate [13]. We found that SG grown $Mn_xSi_{1-x}$ ($x \approx 0.5$) films have two magnetic phases: HT FM phase with $T_C \approx 370$ K and LT FM phase with $T_C \approx 46$ K. At the same time, the anomalous Hall effect (AHE) changes its sign from the positive to negative one at the temperature below 30 K. We explain obtained experimental results by the interplay of two effects: 1) self-organization of polycrystalline film leading to the formation of two layers with strongly differing sizes of crystallites; 2) peculiarities of defect-induced FM ordering in such a system.

**2. Samples and experimental details**

The SG grown $Mn_xSi_{1-x}$ thin films were deposited in Kr atmosphere (~$10^{-2}$ mbar) onto the $Al_2O_3$ (0001) substrates 10x15 mm² in size using the single crystal MnSi target [13]. The substrate temperature during the deposition (340 °C) was the same as for previous DG deposited films, while the deposition rate was higher ($\geq 7$ nm/min). The Rutherford backscattering spectrometry (RBS) was used to determine the film composition and thickness [13]. The film thickness $d$ depends on the distance $L$ to the target; the value $d$ decreases from 270 to 70 nm with the increase of $L$ at the length $\delta L \approx 15$ mm. The Mn content at the same deposited area increases from 0.506 up to 0.517. When the film thickness decreases from 160 to 70 nm ($\delta L \approx 10$ mm), the film composition changes only slightly with $L$ ($x \approx 0.514$-$0.517$). To investigate the effect of the film composition and deposition rate on the magnetic and magneto-transport properties, the as grown sample was cut into seven 2x10 mm$^2$ stripes with different thicknesses. Here we present the results for the most distant from the target samples with slightly changing Mn content $x \approx 0.514$-$0.517$ and different film thickness 70-160 nm.

The structural properties of Mn-Si samples were investigated by X-ray diffraction (XRD) measurements using a Rigaku SmartLab diffractometer. To elucidate the microscopic structure of Mn-Si films, they were further investigated by scanning transmission electron microscopy (STEM) using TITAN 80-300 TEM/STEM instrument (FEI, US) operating at an accelerating voltage of $U$=300 kV, equipped with Cs-probe corrector, high-angle annular dark-field detector (HAADF) (Fischione, US) and energy dispersive X-ray (EDX) microanalysis spectrometer (EDAX, US). Cross-section transmission electron microscopy (TEM) specimens were prepared by the mechanical polishing of sandwiched pieces followed by Ar$^+$ ion-beam milling until perforation in Gatan PIPS (Gatan, US).

## 3. Magnetic and magneto-transport measurements

The temperature dependence of saturation magnetization $M_s(T)$ of three Mn$_x$Si$_{1-x}$ samples 1-3 ($x \approx 0.517$, 0.516 and 0.514) with the thicknesses $d \approx 70$, 90 and 160 nm, respectively, is presented in Fig. 1. The applied field was $\mu_0 H = 1$ T. The obtained data of Fig.1 revealed a presence of two ferromagnetic phases: a HT phase with $T_C \approx 370$ K and a LT phase with $T_C \approx 46$ K. The relative contribution of the LT FM phase clearly increases with the increase of the film thickness. Such behavior is in contrast to that of DG grown Mn$_x$Si$_{1-x}$ films (Fig. 2). When $x \approx 0.52$, the decrease of $M_s(T)$ in the temperature range $T = 10$-$100$ K does not exceed 6% and fits well to the Bloch law [13]. Moreover, the $M_s(T)$ value does not increase significantly with lowering $T$ even in case of HT FM degradation, as observed in DG films with the Mn content $x \geq 0.53$ (Fig. 2, see also [3]).

Fig. 3 shows the magnetization vs. magnetic field $M(H)$ dependence for the sample 1 ($x \approx 0.517$, $d \approx 70$ nm) at $T = 5$, 100 and 300 K. The hysteresis loop opens at temperature below 100 K (see inset in Fig. 3), which is not observed in bulk $\varepsilon$-MnSi single crystal. The magnetization saturates in the magnetic field $\mu_0 H \approx 0.6$ T at low temperature ($T = 5$ K) and then linearly increases like in the case of $\varepsilon$-MnSi single crystal [9, 14].

It is curiously to note that at $T > 46$ K the "surface density" of magnetic moment $J_m/A$ of the HT FM phase (i.e. the total magnetic moment $J_m$ normalized to the film surface $A$; see inset to Fig. 1) does not depend on the film thickness. This fact clearly indicates that the HT FM phase ($T_C \approx 370$ K) is formed only in the interfacial layer directly deposited on the substrate with a fixed thickness, while the LT FM phase ($T_C \approx 46$ K) is formed in the upper layer with a variable thickness. So, the magnetometry data allude to a presence of two FM layers with different thicknesses, magnetic moments and Curie temperatures in our films.

The Hall effect provides rich information on the correlation between magnetic and transport properties of Mn-Si system under investigation. Let us recall that in "ordinary" FM material, the Hall resistance $R_H$ contains two components following the expression [15]:

$$R_H d = \rho_H = \rho_H^n + \rho_H^a, \quad \rho_H^n = R_0 B, \quad \rho_H^a = R_s M, \tag{1}$$

where $\rho_H$ is the total Hall resistivity, $\rho_H^n$ and $\rho_H^a$ are the normal and anomalous components of the Hall resistivity, respectively, $d$ is the thickness of FM material, $R_0$ is the normal Hall effect constant related to the Lorentz force, $B$ is magnetic induction, $R_s \propto (\rho_{xx})^\alpha$ is the anomalous Hall effect (AHE) constant related to the spin-orbit interaction in FM material, $M$ is the magnetization. For a "skew-scattering" driven mechanism of AHE, $\alpha = 1$, and for "intrinsic" and "side-jump" mechanisms of AHE, index $\alpha = 2$ [15]. Usually, at the temperature $T \leq T_C$ and for the magnetic field corresponding to the saturation magnetization, the second term in Eq.(1) dominates, i.e. $\rho_H^n \ll \rho_H^a$. Note that in the case of $\varepsilon$-MnSi single crystal, the third term may also appear in Eq.(1) due to skyrmions formation [16] (so-called component of the topological Hall effect), but in our system we presume that skyrmions are destroyed due to the scattering on the structural and magnetic disorder in the $Mn_xSi_{1-x}$ alloy.

Fig. 4 demonstrates the magnetic field dependence of $\rho_H(B)$, as measured for the sample 1 ($d \approx 70$ nm, $x \approx 0.517$) at the temperature range $T = 5$-$200$ K. One can see that the anomalous component $\rho_H^a$ in the saturation regime (at $B \geq 1$ T) decreases up to 10 times as the temperature decreases from 200 K to 5 K. One can notice that in case of the DG grown film the value of $\rho_H^a$ in the same temperature range is either nearly constant (for $x \approx 0.52$) or increases as the temperature lowers (up to 2 times for $x \approx 0.55$, see [3]). The unusual behavior of $\rho_H(B)$ in the SG grown film can be explained as a partial compensation of the positive Hall emf from the bottom HT FM layer and the negative Hall emf from the upper LT FM layer (see inset to Fig. 4). To justify this explanation we have to suggest that in the upper layer, the effect of LT FM order on the Hall transport is similar to the case of bulk $\varepsilon$-MnSi, where AHE has the negative sign [14, 16]. At the same time, we have to postulate that in the bottom layer, the effect of the HT FM order on the Hall transport is similar to the case of DG films [3], where the AHE of positive sign was reported [3] (the AHE of positive sign is observed also in amorphous $Mn_xSi_{1-x}$ alloys [7, 17]). Evidently, in the two-layer SG grown film a partial compensation of negative and positive contributions $\rho_H(B)$ should be more pronounced at temperatures below the Curie temperature of the LT FM layer ($T_C \approx 46$ K); this compensation becomes more efficient with the film thickness increasing.

The temperature dependence of $\rho_H(T)$ of the thicker sample 2 ($d \approx 90$ nm, $x \approx 0.516$) measured at $B = 1.2$ T is presented in Fig. 5a. One can see that below $T \approx 50$ K the $\rho_H(T)$ function falls down and then changes its sign to the opposite below $T \approx 30$ K. In the temperature range $T < 30$ K, the hysteresis loop $\rho_H(B)$ acquires unusual shape (Fig. 5b). Obviously, this is the result of superposition of two AHE components: the first one is hysteretic and provided by HT FM layer, $\rho_{H1}^a > 0$, while the second one is non-hysteretic and provided by the LT FM layer, $\rho_{H2}^a < 0$. Notice, that due to the larger values of thickness and conductivity of the LT FM layer, its contribution to the Hall resistance is larger in absolute value than that from the HT FM layer (see Eq. 4 below).

The positive sign of $\rho_{H1}^a$ component is not surprising and testifies to a similarity of structural, magnetic and transport properties of the bottom HT FM layer and DG grown $Mn_xSi_{1-x}$ films. The negative sign of $\rho_{H2}^a$ may be attributed to a similarity of the properties of the upper LT FM layer and $\varepsilon$-MnSi, where $\rho_H^a$ is negative [14, 16]. It is also important to note that normal Hall effect in $\varepsilon$-MnSi is positive [14, 16]; therefore, the linear behavior of the $\rho_H(B)$ dependence in fields the $B \geq 0.7$ T corresponds to the hole type of conductivity (see Fig. 5b).

In order to analyze better the results of Hall effect measurements in a two-layer system, we have also studied the temperature dependence of longitudinal resistivity $\rho(T)$ for grown SG films. In Fig.6, the normalized temperature dependences $\rho(T) = \rho_{SG}(T)$ and $\rho(T) = \rho_{DG}(T)$ (taken from [3]) are shown, respectively, for SG and DG grown $Mn_xSi_{1-x}$ films ($d \approx 70$ nm; $x \approx 0.52$), in comparison with $\rho(T) = \rho_{SC}(T)$ for $\varepsilon$-MnSi single crystal (taken from [18]). Note the similarity between $\rho_{SG}(T)$ and $\rho_{SC}(T)$ and its drastic difference from $\rho_{DG}(T)$.

## 4. Structure measurements

The results of XRD measurements of as grown $Mn_xSi_{1-x}/Al_2O_3(0001)$ samples 10x15 mm$^2$ in size (before cutting) are shown in Fig. 7. The angular range $2\theta = 30\text{-}70°$ contains several peaks, which are all attributed to $\varepsilon$-MnSi phase with B20 structure. An additional intense diffraction peak observed at $2\theta = 64.5°$ does not belong to $\varepsilon$-MnSi and could point at (200) plane diffraction of *c*-MnSi phase (similarly to *c*-FeSi phase in Ref. [19]). However, further analysis of XRD rocking curve reveals that this peak is due to a quasi-forbidden reflection (0009) from the Al$_2$O$_3$ substrate and appears as a result of multiple reflections (so-called multi-wave diffraction, see insert to Fig. 7).

The results of the TEM analysis, particularly, low magnification bright field TEM image of Mn$_x$Si$_{1-x}$/Al$_2$O$_3$(0001) is shown in Fig. 8a. Mn$_x$Si$_{1-x}$ film has a columnar microstructure with the lateral grain sizes of about 50 nm. The electron diffraction (ED) study and Fourier analysis of lattice images (not shown) pointed to the B20 type of crystal structure of the MnSi film consistent with XRD data. Dark-field high resolution STEM images of the Mn$_x$Si$_{1-x}$/Al$_2$O$_3$(0001) interface, shown in Fig 8b, revealed the presence of nanometer size crystallite layer near the interface with the thickness of ~10 nm. The grains exhibit equiaxed morphology with the size of ~5 nm. The Fourier analysis of High Resolution TEM images (Fig. 8 c-e) evidences that these crystallites adopt B20 crystal structure of $\varepsilon$-MnSi single crystal.

## 5. Discussion

The results of TEM investigation clearly indicate a two-layer structure in the studied Mn$_x$Si$_{1-x}$ ($x \approx$ 0.51-052) films, apparently due to the peculiarity of the SG growth process. The important difference between the structures of each layer is their grain sizes which are adopted during the growth. The bottom interfacial HT FM layer directly deposited on the substrate is composed of the nanometer size crystallites (~5 nm) and has the fixed thickness (~10 nm), while the upper LT FM layer is practically homogeneous on the nanometer scale and changes its thickness from ~ 60 nm to ~ 150 nm in studied SG grown films. On the basis of this two-layer picture, let us analyze the data of magnetic and transport measurements of Mn$_x$Si$_{1-x}$ films ($x \approx$ 0.51-0.52).

First of all, we have to estimate the value of effective magnetic moment on Mn atom in both layers, suggesting that the density of Mn$_x$Si$_{1-x}$ ($x \approx$ 0.51-0.52) alloy is equal to that of the bulk $\varepsilon$-MnSi single crystal, i.e. $\approx$ 5.82 g/cm$^3$ [20]. The HT FM and LT FM phase contributions to the total magnetization of the film can be found using the simplified Brillouin function fit for $M_s(T)$:

$$M_s(T) = M_s(0)[1-(T/T_C)^n]. \qquad (2)$$

In our case, $n = 1.5$ leads to the best fit of experimental $M_s(T)$ data (Fig. 1). Using Eq. (2), we have found for the samples with $x \approx$ (0.51-0.52) the effective magnetic moments $m$ = (1.3-1.75) $\mu_B$/Mn and (0.43-0.52) $\mu_B$/Mn for for HT and LT FM phase, respectively.

The effective magnetic moment of the bottom HT FM layer in $Mn_xSi_{1-x}$ films ($x \approx 0.51$-$0.52$) grown in the SG significantly exceeds the magnetic moment of MnSi single crystal, $m \approx 0.4\ \mu_B/Mn$ [9]. It is also higher as compared to the case $Mn_xSi_{1-x}$ film grown in DG ($x \approx 0.52$, $T_C \approx 330K$), where the effective magnetic moment is $m \approx 1.1\ \mu_B/Mn$ [3]. These facts do not leave doubt about existence of defect-induced local magnetic moments in the HT FM phase, which are formed due to the same mechanism as in DG grown nonstoichiometric $Mn_xSi_{1-x}$ ($x \approx 0.52$) alloys. The origin of this mechanism, following Ref. [3], consists in the variation of coordination number of Mn atom near the Si vacancy. Due to a strong hybridization between 3d-electron states of Mn and 3(s,p)-electron states of Si this variation leads to the corresponding local redistribution of charge and spin densities near the Si vacancy, which is thereby responsible for the formation of a complex defect with local magnetic moment ~(2.0-3.5)$\mu_B$/Mn and effective (average) magnetic moment ~(1.2-1.75)$\mu_B$/Mn.

The effective magnetic moment of the upper LT FM layer is in good agreement with the magnetic moment of single crystalline $\varepsilon$-MnSi; this fact may be naturally interpreted as an absence of local magnetic moments in the upper LT FM layer. At first glance, this conclusion is surprising, since according to the results of TEM studies and Rutherford backscattering analysis [13] the composition of the film is homogeneous across the film thickness, i.e. LT FM phase contains the same excess amount of Mn atoms as in HT FM phase. Therefore, we have to suggest that most part of Mn containing defects in the upper LT FM layer is in a weak-magnetic or non-magnetic ("magnetically dead") configuration. Following Ref. [3, 4], as an example of such the configuration we can imagine an interstitial Mn atom introduced into the MnSi matrix. The calculated magnetic moment on this Mn atom is extremely small (~0.09$\mu_B$/Mn) and the effective (average) magnetic moment is ~0.34$\mu_B$/Mn for $Mn_xSi_{1-x}$ ($x \approx 0.52$) film.

To explain magnetic data we suppose that due to the specificity of the SG method the Si vacancies mainly arise in the lower layer of the film. The nanocrystallite boundaries in this layer form a vast network; they eventually can work as the gettering regions for Si vacancies and, consequently, for local magnetic moments on these vacancies. So, following our supposal, nanocrystallite boundaries play the key role in the magnetic properties of HT FM layer, acting as a magnetic envelope of the nanometer scale non-magnetic crystallite. Early in Ref. [12] in the frame of the spin-fluctuation model of FM ordering, we have analyzed the role of dimension effects in granular dilute Si-Mn alloys. We considered the precipitate nanoparticles of $SiMn_{1.7}$ type in the Si matrix and estimated variation of the Curie temperature as a function of the shape and size of these precipitates. Similar analysis can be effectuated for the case of $Mn_xSi_{1-x}$ ($x \sim 0.5$) alloys. For a spherical crystallite of weak itinerant FM with the small radius $r_0 \ll \zeta$, where $\zeta$ is FM correlation length, encircled by an envelope with defect–induced local magnetic moments $S$ having the surface density $\sigma_0 \ll a^{-2}$, where $a$ is the lattice parameter, we can roughly estimate the Curie temperature $T_C$ as

$$k_B T_C \sim JS(W/v_F Q_{SF})^{1/2} (a/r_0)^{1/2} (\sigma_0 a^2)^{1/2}. \qquad (3)$$

Here $J$ is exchange interaction potential between the local moment on the defect and itinerant electron spin, $W$ is itinerant electron bandwidth, $v_F$ is the Fermi velocity, $Q_{SF}$ is spin-fluctuation cutoff wave vector. At $JS \sim 0.1$ eV, $W/v_F Q_{SF} \sim 10$, $a/r_0 \sim 10^{-1}$, $\sigma_0 a^2 \sim 10^{-1} \div 10^{-2}$ we have $T_C \sim$ 100÷400 K that is not far from above obtained experimental results.

Let us now consider the transport data. As the temperature decreases from 300 to 5 K, the $\rho_{DG}(T)$ curve in Fig.6 demonstrates a relatively slow (about 1.3 times) temperature decreasing. It was also shown in Ref. [3] that, contrary to the case of $\varepsilon$-MnSi single crystal, for the DG films the carrier mobility strongly increases (about fifteen times at 60 K), but the carrier concentration drastically decreases (about twenty five times at 100 K). Thus, $\rho_{DG}(T)$ behavior is driven by an interplay of these two effects and as a result, the value $\rho_{DG}(T)$ for $Mn_xSi_{1-x}$ ($x \approx 0.52$) film below ~ 40 K significantly exceeds $\rho_{SC}(T)$ for $\varepsilon$-MnSi, where $\rho_{SC}(T)$ falls down dramatically [18].

The physical origin of this remarkable phenomenon of simultaneous increase of carrier mobility and decrease of carrier concentration at the doping of single crystal $\varepsilon$-MnSi with additional Mn atoms is not yet clear. A possible (but certainly open to discussions) reason qualitatively explaining experimental data has been proposed in Ref. [3]. It presumes that: 1) the Mn doping induces the carrier localization on the defect center (e.g., the above discussed Si vacancy) in the MnSi matrix; 2) this doping also destroys collective (Kondo or spin-polaron type) resonance, probably existing in $\varepsilon$-MnSi single crystal. The combination of these two effects obviously leads to the simultaneous decrease of carrier concentration and the increase of carrier mobility, if we suggest that the additional carrier mobility decrease due to the carrier scattering on the defects is small compared to the carrier scattering on the collective resonance.

The temperature resistivity dependence $\rho_{SG}(T)$ in the high temperature region (above $T$ ~250 K) has almost the same character as $\rho_{DG}(T)$, but differs from it at low temperatures (see Fig. 6). Between $T = 250$K and 40K, the $\rho_{SG}(T)$ function decreases almost 1.6 times more than $\rho_{DG}(T)$; below $T \approx 40$ K, the $\rho_{SG}(T)$ function falls down similar $\rho_{SC}(T)$ in the case of $\varepsilon$-MnSi single crystal (Fig. 6), although not so drastically. Obviously, that extraction of a serious physical information from the direct comparison of $\rho_{SG}(T)$ and $\rho_{DG}(T)$ is hampered, since the SG film has a two-layer structure, but the DG film is homogeneous. The problem is to estimate correctly the contribution of each layer to $\rho_{SG}(T)$. Assuming that conductivities of both layers are of the same order, we can roughly suggest that for thick films ($d \sim d_2 \gg d_1$) the function $\rho_{SG}(T)$, which has almost the same character as for $\varepsilon$-MnSi single crystal (Fig. 6), mainly corresponds to the temperature dependence of the resistivity of the upper LT FM layer. Thus, at least for qualitative purposes we may fancy the upper LT FM layer as the $\varepsilon$-MnSi single crystal with non-magnetic electro-neutral defect centers and not completely destroyed collective resonance. Unfortunately, it is difficult to conclude definitely about the internal structure of the lower HT FM layer if one takes into account only $\rho_{SG}(T)$ data. We can only speculate that the material of this layer is similar to the one of DG film.

We are able to obtain additional information on the physical properties of the LT FM and HT FM phases analyzing the magneto-transport data for the SG film. If we present this film as two parallel conducting layers (see inset to the Fig. 4) the effective Hall resistivity can be written as

$$R_H = \frac{\rho_{H2}\sigma_2^2 d_2 + \rho_{H1}\sigma_1^2 d_1}{(\sigma_1 d_1 + \sigma_2 d_2)^2}, \qquad (4)$$

where the indices "1" and "2" correspond to the lower (HT FM) and upper (LT FM) layer, respectively. From Eq. (4) it is seen that in thick films ($d \sim d_2 \gg d_1$) the change of the Hall effect sign is possible when temperature decreases below upper layer Curie temperature ($T_{C2} \approx 46$ K) and the negative anomalous component of the Hall effect ($\rho_{H2}^a < 0$) in this layer starts to play a dominant role due to its similarity to the case of bulk $\varepsilon$-MnSi [14, 16].

The ratio between the conductivities of two layers $\sigma_2/\sigma_1$ can be found using following assumptions: 1) the AHE resistivity of lower layer at $T < 200$ K is the same as for DG film [3], i.e. $\rho_{H1}^a \approx +3.5 \cdot 10^{-6}$ $\Omega \cdot$cm; 2) the AHE resistivity of upper layer at $T = (25-40)$ K is the same as for $\varepsilon$-MnSi single crystal [14, 16], i.e. $\rho_{H2}^a \approx -(0.1-0.2) \cdot 10^{-6}$ $\Omega \cdot$cm; 3) the sign of the Hall effect changes to the opposite at the thickness $d = d_2 + d_1 = (70-90)$ nm (Figs. 4 and 5). Substituting these data in Eq.(4) we obtain the ratio $\sigma_2/\sigma_1 \approx 2$. In other words, in spite of significant decrease of the nano-crystallites size in the bottom layer compared to that in the upper layer, the conductivity of bottom layer at low temperature does not significantly decrease. Probably, we observe here the effect of partial compensation of two effects (carrier concentration decrease and carrier mobility increase) having the same physical origin as in above discussed case of DG thin film [3].

**6. Conclusions**

In this work, we present for the first time the results of comparative study of magnetic and transport properties of nonstoichiometric $Mn_xSi_{1-x}$ ($x \approx 0.51-0.52$) films grown by the PLD technique onto the single crystal $Al_2O_3(0001)$ substrates at $T = 340°C$ using SG and DG method.

The key point of SG approach is the using Kr as a scattering gas which results in the lower energy of deposited atoms. At the same time, the average deposition rate in SG is much higher ($\geq 7$ nm/min) than in DG. The SG grown $Mn_xSi_{1-x}$ films on the rectangular substrate 10x15 mm$^2$ in size possess slightly varying composition ($x$=0.506-0.517) and large variation in thickness ($d$=270-70 nm) depending on the distance from the Mn-Si target.

X-ray diffraction analysis reveals that textured $\varepsilon$-MnSi-like phase with the B20-type crystal structure dominates in both SG and DG type of films. While the $\varepsilon$-MnSi single crystal has the Curie temperature $T_C \approx 30$ K [8, 9], the studied $Mn_xSi_{1-x}$ films at $x \approx 0.52$ exhibit HT FM with $T_C > 300$ K accompanied by the manifestation of the positive sign of AHE. For SG grown $Mn_xSi_{1-x}$ films, it is found that at low temperature the essential contribution to the magnetization is given by LT FM phase with $T_C \approx 46$ K; at the same time, AHE changes the sign from the positive to negative at $T \leq 30$K and film thickness $d \geq 90$ nm.

We explain these results as the manifestation of self-organizing effect in the SG polycrystalline $Mn_xSi_{1-x}$ film, i.e. the formation of two layers with significantly different thickness and grain size, leading to the opposite sign contributions in to AHE. The bottom interface layer adjacent to $Al_2O_3$(0001) substrate is ~10 nm in thickness with $T_C \approx 370$ K and consists of small (~ 5 nm) rounded grains. The top layer ~60 – 260 nm in thickness with a columnar grain structure ~50 nm in lateral size represents LT phase, which exhibits negative AHE similar to that in the $\varepsilon$-MnSi single crystal [14, 16].

Finally, we discuss obtained experimental results in terms of the model of defect-induced FM order with effective exchange coupling strongly affected by spin fluctuations [12] taking into account the structure peculiarities of studied films. We argue that the observed HT FM of nonstoichiometric $Mn_xSi_{1-x}$ alloys strongly depends on the type of defects ("magnetically active" Si vacancies vs. "magnetically dead" interstitial Mn atom) as well as on the size of crystal grains which interfaces acting as the gettering regions for Si vacancies.

**Acknowledgements**

The work was partly supported by the RFBR (grant Nos. 14-07-91332, 14-07-00688, 14-47-03605, 14-22-01063, 15-29-01171, 13-07-12087, 13-07-00477), NBICS Center of the Kurchatov Institute and MIPT Center of Collective Usage with financial support from the Ministry of Education and Science of the Russian Federation (Grant No. RFMEFI59414X0009). The work at HZDR is financially supported by DFG (ZH 225/6-1). A.S.S. acknowledges the financial support of DAAD-MSU program "Vladimir Vernadsky".

# Figure captions

Fig. 1. Temperature dependence of saturation magnetization $M_s$ for $Mn_xSi_{1-x}$ films with different thickness and close Mn content ($x \approx 0.516$) grown in shadow geometry. The insert shows the temperature dependence of magnetic moment $J_m$ normalized by film square $A$. (For sample with $d = 70$ nm the $J_m(T)/A$ curve practically coincides with one for sample with $d = 90$ nm and is not shown on the insert).
Solid lines are fitting dependencies of $M_s(T)$ with using equation (2).

Fig. 2. Temperature dependence of saturation magnetization $M_s$ for $Mn_xSi_{1-x}$ films with $x \approx 0.52$ и 0.53 ($d \approx 70$ nm) grown at "direct" geometry by PLD. Solid line is calculated dependence of $M_s(T)$ from [3].

Fig. 3. Magnetization versus magnetic field for SG grown sample 1 ($d \approx 70$ nm; $x \approx 0.517$) at different temperatures. The insert shows $M(H)$ dependences in an enlarged scale.

Fig. 4. Resistivity of the Hall effect versus magnetic field for SG grown sample 1 ($d \approx 70$ nm; $x \approx 0.517$) at different temperatures. The inset shows the cross-section of $Mn_xSi_{1-x}/Al_2O_3$ structure.

Fig. 5. (a) Temperature dependence of the Hall resistivity for SG grown sample 2 ($d \approx 90$ nm; $x \approx 0.516$) measured at $B = 1.2$ T. (b) Resistivity of the Hall effect versus magnetic field for sample 2 at $T = 9K$.

Fig. 6. Normalized temperature dependence of the longitudinal resistivity $\rho(T)$ for DG (curve 1) and SG (curve 2) grown $Mn_xSi_{1-x}$ films ($d \approx 70$ nm; $x \approx 0.52$) in comparison with $\rho(T)$ for $\varepsilon$-MnSi (taken from [18]).

Fig. 7. The results of X-ray diffractometry for SG $Si_{1-x}Mn_x/Al_2O_3(0001)$ structure. Insert shows quasi-forbidden reflection (0009) from $Al_2O_3$ substrate.

Fig. 8. The cross-section images and the study of crystal structure of SG $Mn_xSi_{1-x}/Al_2O_3(0001)$ sample: (a)-BF image of the film. (b)- HAADF STEM image of the $Mn_xSi_{1-x}/Al_2O_3(0001)$ interface. (c)- HRTEM image of the $Mn_xSi_{1-x}/Al_2O_3(0001)$ interface area. Several grains studied by Fourier analysis are marked by red rectangles. (d)-enlarged HREM image of one grain. (e)-Fourier spectra of that grain which matches to B20 MnSi crystal structure in [102] zone axis.

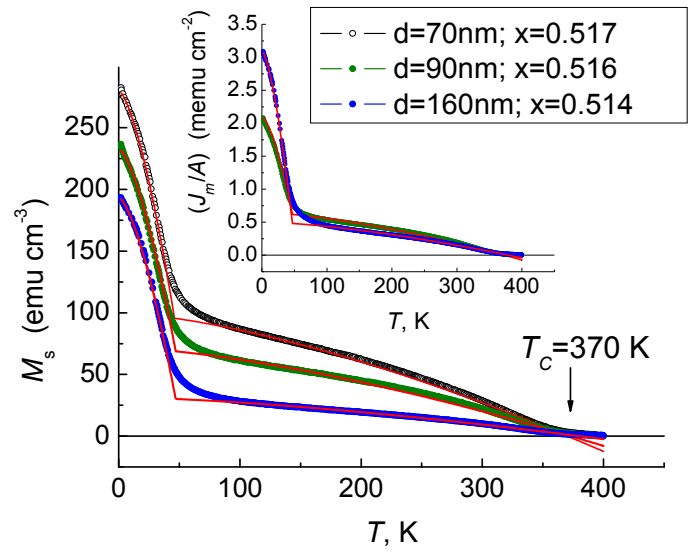

**Fig. 1.**

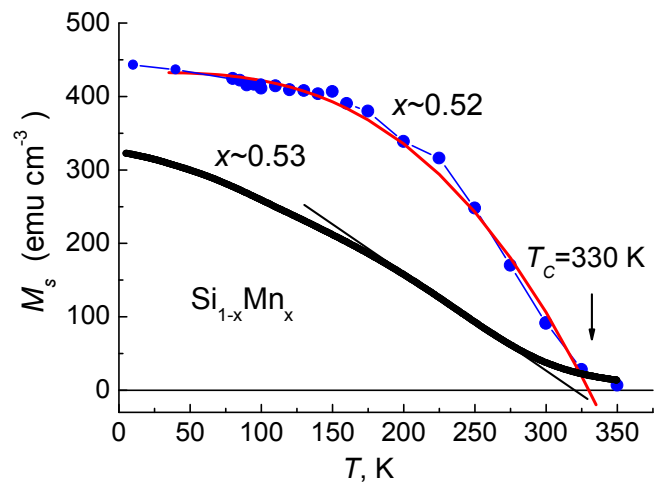

**Fig. 2.**

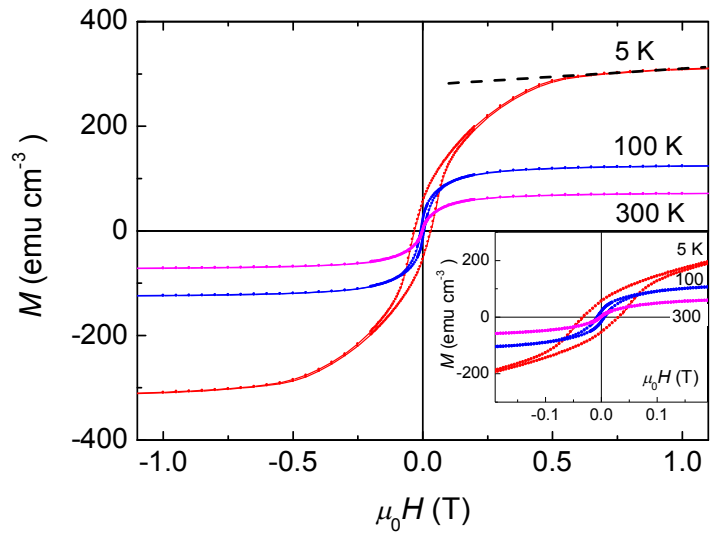

**Fig. 3.**

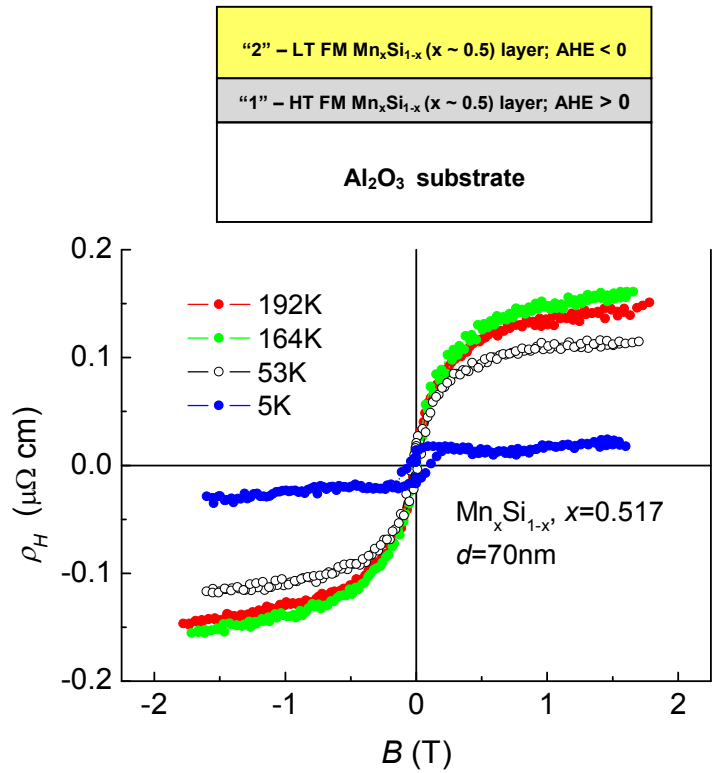

**Fig. 4.**

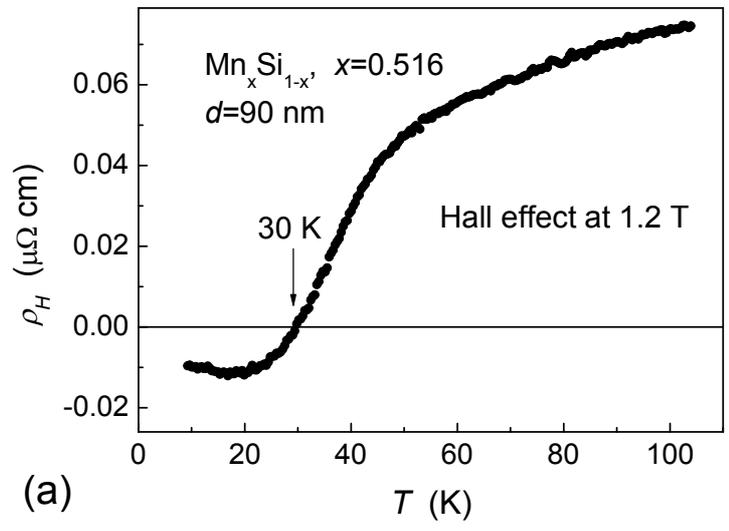

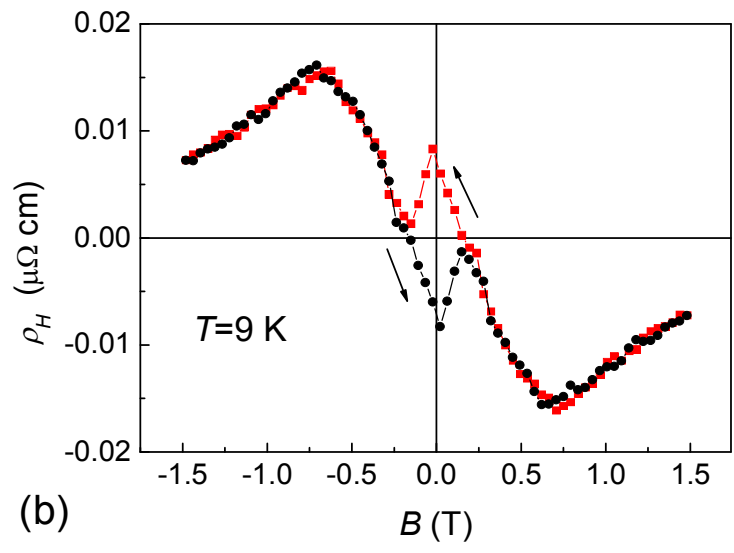

**Fig. 5.**

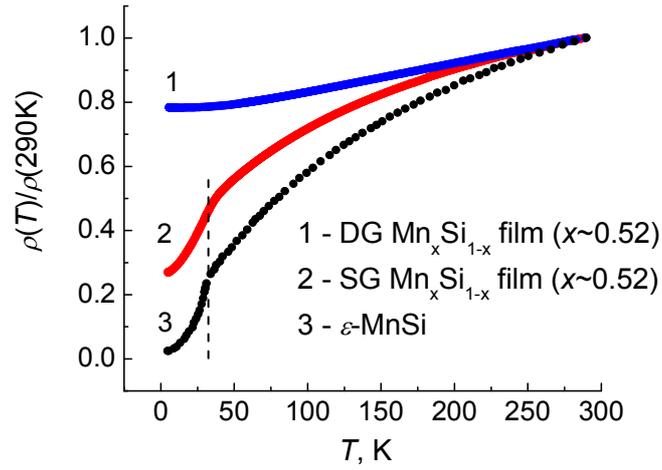

**Fig. 6.**

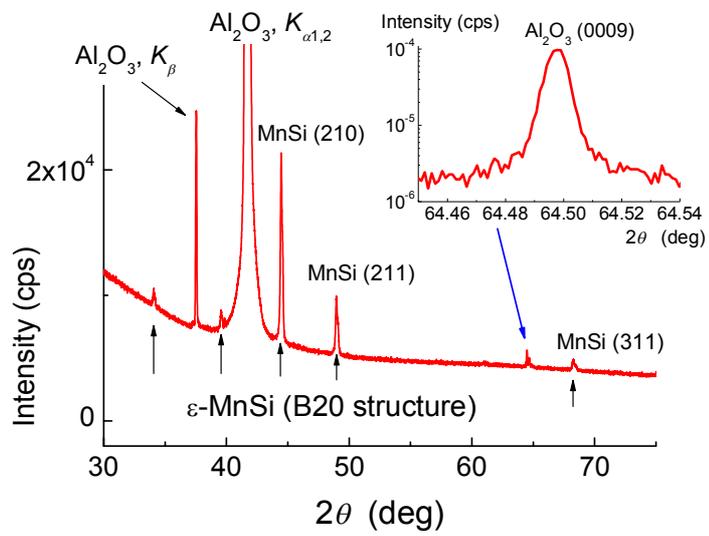

**Fig. 7.**

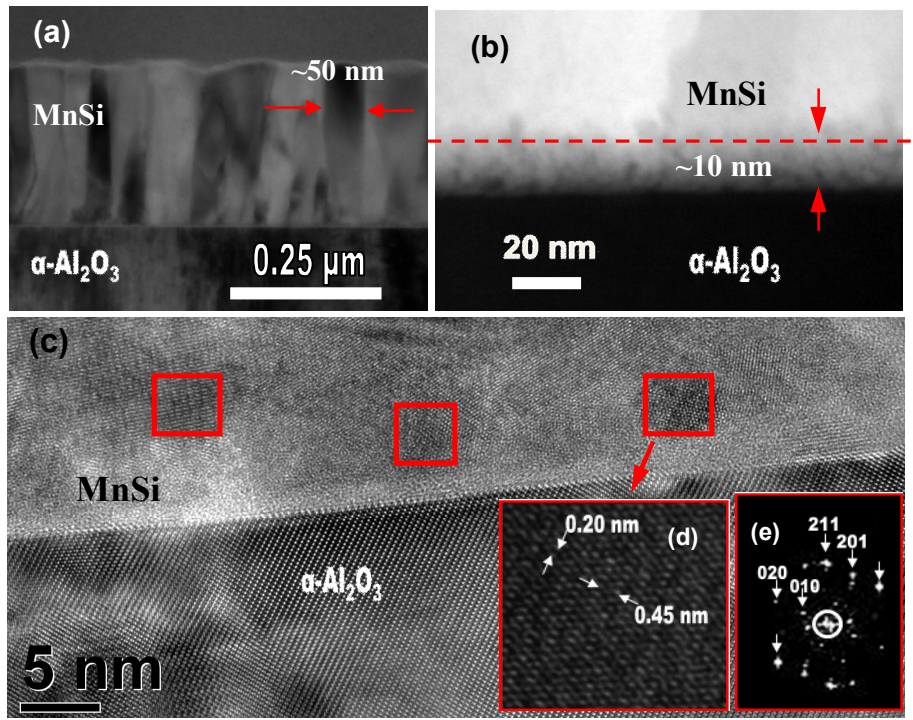

**Fig.8.**